\documentclass[10pt,aps,prb,twocolumn,groupedaddress,longbibliography]{revtex4-1}
\usepackage{amsmath}
\usepackage{amsfonts}
\usepackage{amssymb}
\usepackage[utf8]{inputenc}
\usepackage{graphicx}
\usepackage{tabularx}
\usepackage{color}
\usepackage{caption}

\begin{document}

%%%%%%%%%%%%%%%%%%%%%%%%% Title %%%%%%%%%%%%%%%%%%%%%%%%%%%%%%%%%%%%%%%
\title{Improved CPS and CBS Extrapolation of PNO-CCSD(T) Energies: The MOBH35 and ISOL24 Data Sets}

\author{Kesha~Sorathia}
\author{Damyan~Frantzov}
\author{David~P.~Tew}
\email{david.tew@chem.ox.ac.uk}
\affiliation{
University of Oxford, South Parks Road, Oxford, OX1 3QZ, UK
}

\date{\today}

\begin{abstract}
Computation of heats of reaction of large molecules is now feasible using domain-based PNO-CCSD(T) theory. However, 
to obtain agreement within 1~kcal/mol of experiment, it is necessary to eliminate basis set incompleteness
error, which comprises of both the AO basis set error and the PNO truncation error. Our investigation into the 
convergence to the canonical limit of PNO-CCSD(T) energies with PNO truncation threshold $T$ shows that errors follow the model 
$E(T) = E + A T^{1/2}$. Therefore, PNO truncation errors can be eliminated using a simple two-point CPS 
extrapolation to the canonical limit, so that subsequent CBS extrapolation is not limited by residual PNO truncation error.
Using the ISOL24 and MOBH35 data sets, we find that PNO truncation errors are larger for molecules with 
significant static correlation, and that it is necessary to use very tight thresholds of $T=10^{-8}$ to 
ensure errors do not exceed 1~kcal/mol. We present a lower-cost extrapolation scheme that uses information from 
small basis sets to estimate PNO truncation errors for larger basis sets. In this way the canonical limit 
of CCSD(T) calculations on large molecules with large basis sets can be reliably estimated in a practical way. Using
this approach, we report complete basis set limit CCSD(T) reaction energies for the full ISOL24 and MOBH35 data sets.
\end{abstract}

\maketitle

\section{Introduction}

Heats of reaction and activation enthalpies computed using the
coupled cluster singles, doubles and perturbative triples method, CCSD(T),\cite{Raghavachari:CPL157-479}
are often accurate to within 1~kcal/mol of experimentally derived values.\cite{mest}
Even though CCSD(T) is based on a single Hartree--Fock (HF) reference wavefunction,
the correlation treatment is complete to fourth-order in perturbation theory and orbital relaxation is accounted
for self-consistently through the singles excitations. CCSD(T) energies are frequently found to be
accurate for systems where HF energies are poor, for example in some transition metal
complexes, even though they exhibit large T1-diagnostics.\cite{LiManni:JCTC15-1492,Giner:JCTC14-6240,Tew:JCP145-074103} 

A great deal of effort has been spent on reducing the high computational 
cost of CCSD(T) to increase the size of system that can be modelled, for example through
massively parallel implementations,\cite{Rendell:CPL194-84,Valiev:CPC181-1477} fragmentation 
methods\cite{Friedrich:JCTC5-287,Kjorgaard:WIRES7-6,Rolik:JCP139-094105,Nagy:JCTC15-5275,Nagy:JCTC14-4193,Usvyat:WIRES8-e1357} 
and local correlation methods.\cite{Pulay:CPL100-151,Saebo:ARPC44-213,Werner:JCP135-144116,Neese:JCP131-064103,Maurer:JCP138-014101}
Local approximations exploit the short-range nature of electron correlation 
to reduce the scaling from $\mathcal{O}(n^7)$ for CCSD(T) to sub-quadratic in system size $n$,
such that calculations on very large molecules are possible,\cite{Franzke:JCTC} albeit with some loss of accuracy arising
from the neglected contributions.\cite{Sylvetsky:JCTC16-3641}

This article is concerned with the domain-based pair natural orbital (PNO) approach to local correlation,\cite{Riplinger:JCP139-034106,Werner2017,Tew:JCTC15-6597} which is particularly effective and
has found widespread application in both single-reference\cite{Neese:JCP131-064103,Hansen:JCP135-214102,Saitow:JCP146-164105,Calbo:JCC38-1869,Fiedler:JCTC13-6023} and multireference\cite{Guo:JCP144-094111,Menezes:JCP145-124115,Uemura:JCP158-154110,Saitow:JCP157-084101} correlation theories.
In the PNO-CCSD(T) approach, amplitudes from MP2 theory are used to form natural orbitals for each pair of localised 
occupied orbitals, and the full CCSD(T) correlation treatment is performed in a truncated subset of these PNOs. The
size of the subset and the corresponding error incurred is controlled through a user-defined threshold $T$,
which determines the maximum occupation number of the retained PNOs.

The increased overhead of pairwise integral transformation and non-orthogonality of PNOs between pairs is outweighed
by the compression of the T2 amplitude space from $\mathcal{O}(n^4)$ to $\mathcal{O}(n)$ and the associated savings
in evaluating the amplitude working equations. The integral transformation cost is also reduced to $\mathcal{O}(n)$
if PNOs are confined to domains of projected atomic orbitals (PAOs) and if local density fitting is employed.
Domain-based PNO-CCSD(T) has been implemented in the 
Turbomole,\cite{Helmich:JCP135-214106,Helmich:JCP139-084114,Tew:JCP135-074107,Tew:IJQC113-224,Hattig:JCP136-204105,Schmitz:MP111-2463,Schmitz:PCCP16-22167,Schmitz:JCP145-234107,Frank:JCP148-134102,Tew:JCTC15-6597,InC-Tew:2021} 
Orca\cite{Neese:JCP131-064103,Hansen:JCP135-214102,Riplinger:JCP139-034106,Riplinger:JCP139-134101,Pinski:JCP143-034108,Riplinger:JCP144-024109,Pavosevic:JCP144-144109,Pavosevic:JCP146-174108} and 
Molpro\cite{Werner:JCTC11-484,Ma:JCTC11-5291,Ma:JCTC11-5291,Schwilk:JCTC13-3650,Ma:JCTC13-4871,Ma:JCTC14-198,Krause:JCTC15-987,Ma:JCTC16-3135,Ma:JCTC17-902} program packages and is
increasingly being used in studies of chemical stability and reactivity.

Martin and co-workers have recently reported numerical studies that assess the accuracy of PNO-CCSD(T) against
canonical CCSD(T) in the context of metal-organic chemistry.\cite{Semidalas:JCTC18-883}
They find that for systems where 
there is moderate static correlation the PNO truncation error can be several kcal/mol when using
default thresholds of $T=10^{-6.5}$ or $T=10^{-7}$. By tightening the PNO threshold
the canonical result is recovered, but errors under 1~kcal/mol required very tight thresholds of 
$T=10^{-8}$. Sandler \emph{et al} also report sizeable PNO truncation errors for reaction barriers for open and
closed-shell organic reactions when using default settings.\cite{Sandler:JPCA125-1553}

We have previously studied the interdependence of the PNO truncation error and AO basis set error on weakly correlated
systems at the level of MP2 theory.\cite{Sorathia:JCP153-174112} 
The total basis set error is the sum of the intrinsic basis set error due to the chosen AO basis and the 
basis set error made due to the PNO truncation. The intrinsic basis set error affects both the HF and
correlation energies, whereas the PNO truncation error only affects the correlation energy.
For quadruple-zeta basis sets and PNO thresholds of $T=10^{-7}$, we found that the PNO truncation error is 
commensurate with the intrinsic AO basis set error in the correlation energy.
In the cases where the PNO error is dominant, increasing the basis size exhibits a false convergence and
basis extrapolation fails to recover the complete basis set limit.
To reliably apply basis set extrapolation to approach the complete basis set limit it is necessary to use
energies that are closely converged to the canonical values, that is, the limit of a complete PNO space (CPS). 

Care must therefore be taken to control the PNO truncation error when using
PNO methods to accelerate calculation of molecular energies, particularly 
for systems with moderate static correlation or when using large basis sets.
Although simply tightening the PNO threshold in princple guarantees that the canonical result is recovered, the
costs increase by a factor of around 2-3 every ten-fold reduction in $T$. One alternative is to exploit the
systematic reduction in the PNO truncation error and use a series of calculations with decreasing $T$
to extrapolate to the CPS limit, that is, to the canonical result. In this paper we provide 
detailed analysis of CPS extrapolation and give recommendations for best practice.

Altun \emph{et al} explored numerical fits for the behaviour of the PNO truncation error with threshold $T$ and 
proposed the error model\cite{Altun:JCTC16-6142}
\begin{align}
E(T) &= E + B (\log_{10} T)^{\beta} \label{eq:e1} 
\end{align}
$E$ is the energy of the canonical calculation without PNO truncation and $E(T)$ is the energy obtained
using a PNO theshold of $T$, which is typically in the range $10^{-5}$--$10^{-9}$.
This error model does not fit any of our data. Altun \emph{et al}, however, did not use this error model for 
extrapolation, but instead used the the general two-point extrapolation formula
\begin{align}
E &\approx E(T_1) + F ( E(T_2) - E(T_1) )
\end{align}
This approach does not specify an error model, rather the factor $F$ is determined for a chosen pair of thesholds through 
fitting to data. They recommend $F=1.5$ for (6,7) and (7,8) extrapolation, independent of basis set, where (6,7) denotes 
extrapolation with $T_1=10^{-6}$ and $T_2=10^{-7}$.

In simultaneneous work,\cite{Sorathia:JCP153-174112} we proposed an error model motived by the
observation that the energy is proportional to the amplitudes and that the largest discarded amplitude
is proportional to the square root of the PNO truncation threshold $T$. 
\begin{align}
E(T) &= E + A T^{\alpha} \label{eq:e2} 
\end{align}
The exponent $\alpha$ is close to $0.5$ but is allowed to vary with molecule and basis set because the 
converged amplitudes differ from the approximate semi-canonical local MP2 amplitudes used to define the PNO space. 
We demonstrated that the resulting three-point extrapolation scheme applied to MP2 energies reduces the
PNO truncation error in reaction energies equivalent to reducing the tightest PNO threshold by a factor of 50,
essentially eliminating the PNO trunction error without requiring expensive calculations with very tight PNO thresholds.
The three-point extrapolation formula using a sequence of thresholds $T_1 > T_2 > T_3$ is
\begin{align}
E \approx \tilde E = \frac{ E(T_1) E(T_3) - E(T_2)^2 } { E(T_1) + E(T_3) - 2E(T_2)}
\end{align}
Our initial investigations of three-point extrapolation for PNO-CCSD and PNO-CCSD(T) energies, however, was not successful.
We find that the convergence of PNO-CCSD energies with PNO threshold does not fit the error
model used for MP2, due to the differing convergence rates of the ring and ladder terms in the CCSD amplitude
equations. By fixing $\alpha$ to the ideal value of 0.5, a two-point extrapolation formula can be applied.
We find that this approach reduces the PNO truncation error by an amount equivalent to a 10-fold reduction in $T$
for PNO-CCSD(T) energies. Using this approach,
CCSD(T) basis set limit correlation energies of systems with moderate static correlation can be computed
using PNO-CCSD(T) theory without incurring the high cost of very tight PNO thresholds.

Our extrapolation method is operationally very close to that of Altun \emph{et al}. For a given $\alpha$,
two-point extrapolation using our error model results in
\begin{align}
E &\approx  \tilde E = \frac{ T_2^{\alpha} E(T_1) - T_1^{\alpha} E(T_2) }{T_2^{\alpha} - T_1^{\alpha}} 
\label{eq:t2}
\end{align}
which has the equivalent Schwenke\cite{Schwenke:JCP122-014107} form
\begin{align}
\label{eq:t2f}
E  \approx E(T_1) + F ( E(T_2) - E(T_1) ), \quad
F = \frac{T_1^{\alpha}}{T_1^{\alpha} - T_2^{\alpha}} 
\end{align}
By chosing the factor $F=1.5$ for (6,7) and (7,8) extrapolation Altun \emph{et al} are in fact assuming the polynomial error model with $\alpha=0.4771$. A proper understanding of the underlying error model makes it possible to apply the extrapolation using
different choices of PNO threshold, such as (6.5,7), where $F$ becomes 2.366.

In this paper we report our analysis of the PNO truncation errors in PNO-CCSD(T) theory and make recommendations
for reliably extrapolating to the CPS limit to estimate the canonical CCSD(T) results. We use two data sets, the
ISOL24 set of Huenerbein \emph{et al}\cite{Huenerbein:PCCP12-6940} and the MOBH35 set of Iron and Janes.\cite{Iron:JPCA123-3761}
The ISOL24 data set contains relatively weakly correlated systems of up to 81 atoms, and is challenging for PNO methods
because it compares energies of isomers of organic molecules with very different chemical connectivity, spatial arrangements
and long-range dispersion interactions, negating fortuitous error cancellation of local approximations. Werner and Hansen
have very recently reported basis set limit isomerisation energies computed using PNO-LCCSD(T)-F12b theory,\cite{Werner:JCTC} 
which serve as a useful reference point for this work.
The MOBH35 set of metal-organic barrier heights is also challenging for PNO methods since it contains systems with
significant static correlation. The MOBH35 set was used by Semidalas and Martin\cite{Semidalas:JCTC18-883} 
to highlight the slow convergence of the reaction barriers with PNO threshold and larger than expected 
differences in values obtained with different implementations.

\section{Computational details}

All calculations are performed using the Turbomole program package.\cite{Franzke:JCTC}
The structures of the ISOL24 set were taken from the supporting information of Ref~\onlinecite{Huenerbein:PCCP12-6940}.
We use the cc-pVDZ, cc-pVTZ and cc-pVQZ basis sets\cite{Dunning:JCP90-1007} for the PNO-CCSD(T) calculations of these molecules, 
which avoids the problem of internal basis set superposition errors for extended systems.

The structures for the MOBH35 test set were taken from the supporting information of 
Ref~\onlinecite{Semidalas:JCTC18-883}, where the transition state structures for reactions 11 and 12, 
and all species of reaction 14 are modified from the original database, as recommended by Dohm \emph{et al}.\cite{Dohm:JCTC16-2002}
We use the def2-SVP, def2-TZVPP and def2-QZVPP\cite{Weigend:PCCP7-3297} for PNO-CCSD(T) calculations of the MOBH35 set, which
enables direct comparison to earlier work. For molecules containing second- and third-row transition metal atoms,
the Stuttgart relativistic effective core potentials are used.\cite{Dolg:CR112-403} 

For all molecules, Hartree--Fock calculations were performed using the \verb;dscf; program,\cite{HaeserJCC10-104} 
which does not employ the density fitting approximation for the Coulomb integrals. Care was required for reactant 16 
of the MOBH35 set, which converges to the incorrect state if the default extended H\"uckel orbital guess is applied.
The PNO-CCSD(T) calculations
were performed using the \verb;pnoccsd; program in Turbomole V7.7. The Coulomb integrals in PNO methods are approximated
using density fitting and the corresponding Coulomb auxiliary basis sets\cite{Weigend:PCCP8-1057,Weigend:JCP116-3175} 
are used in all cases. 

The domain-based PNO-CCSD(T) implementation in Turbomole uses principal domain theory,\cite{Tew:JCTC15-6597} 
where PAO domains are selected on the basis of an approximate MP2 density in an analogous way to the PNOs themselves.
The approximate MP2 denisty is formed in the basis of orbital specific virtuals neglecting off-diagonal Fock matrix
elements in the occupied space,\cite{Yang:JCP134-044123,Schmitz:MP111-2463} using an OSV truncation threshold linked to the PNO threshold. 
The CCSD ampliutde equations are solved in the basis of retained PNOs and in this work we do not apply weak-pair 
approximations,\cite{Masur:JCP139-164116,Schutz:JCP140-244107} 
since these add additional uncertainty that complicate the analysis of the PNO truncation error. Suppression of the 
weak-pair approximation is acheived using the keyword \verb;multilevel off; in the \verb;$pnoccsd; data group. 
The (T) energy is computed in the basis of
triple natural orbitals\cite{Riplinger:JCP139-134101} using Laplace integration,\cite{Schmitz:JCP145-234107} 
and we use a convergence threshold of 0.01 to determine the Laplace grid. All energies include a correction term that
estimates the energy contribution from discarded pairs and PNOs at the level of MP2 theory, neglecting Fock coupling terms.

One computational bottleneck in PNO methods is the storgage of density fitting intermediates $(Q|ab)$, which are unique to
every pair $ij$ and are required for the ladder terms in the CCSD equations. Despite the fact that the auxiliary functions $Q$ 
are restricted to a pair domain in local density fitting, for large basis sets, tight PNO thresholds and tight density fitting
thresholds, the domain of functions $Q$ and PNOs $a$ is sufficiently large that the required disk space exceeds 1Tb. 
We therefore implemented the possibility to compute the integrals $(ab|cd)$ and $(ab|ck)$ directly, without storing 
the three-index intermediates. This is activated by using the keyword \verb;direct;.

Canonical CCSD(T) calculations were computed using the \verb;ccsdf12; module of the Turbomole package, using 
density fitting for all integrals to ensure that the canonical energies exactly correspond to the CPS
limit of the PNO-CCSD(T) implementation. This is activated using the \verb;risingles; and \verb;riladder; keywords
of the \verb;$ricc2; data group. We were able to compute canonical CCSD(T) energies for the molecules in
reactions 3, 4, 6, 7, 14, 15, 16, 21, 26, 27 and 30-35 using the def2-SVP and def2-TZVPP basis sets.
We denote this subset as MOBH16. We were able to compute the canonical CCSD(T)/cc-pVDZ and CCSD(T)/cc-pVTZ 
energies for all isomer pairs except 1, 4, 6, 7, 16 and 24. We denote this subset as ISOL18.

Where timings are reported, these are performed on a single Intel(R) Xeon(R) Gold 6248R CPU @ 3.00GHz node with
48 cores, 380 Gb RAM and 1.8Tb SSD.

\section{Extrapolation to the CPS limit}

\subsection{Error model $E(T)  = E + A T^\alpha$}

In our previous work, we showed that the error model $E(T)  = E + A T^\alpha$ is very successful for PNO-MP2. 
The work of Altun \emph{et al} indicates that this error model with $\alpha \sim 0.5$ should also be good approximation
for PNO-CCSD(T). The first questions we address in this work are: a) To what extent does this error model fit the PNO truncation
error for coupled-cluster energies?; and b) To what extent does $\alpha$ depend on the molecule and correlation method?

\begin{figure}[!tbp]
   \begin{center}
   \includegraphics[width=0.95\columnwidth]{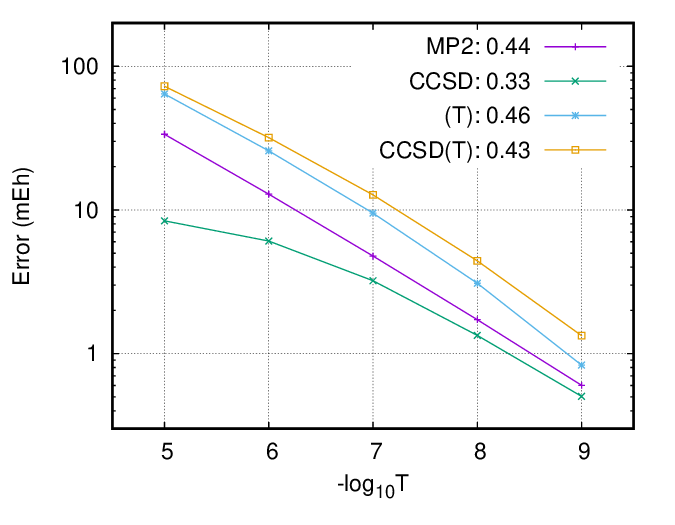}
   \end{center}
\caption{\label{fig:example} PNO truncation errors for educt number 20 of the ISOL24 set using a cc-pVTZ basis. The value of
$\alpha$ in a best fit to $\log_{10}(E(T)-E) = \log_{10} A + \alpha \log_{10} T$ is included in the legend.}
\end{figure}

In Fig \ref{fig:example} we plot the PNO truncation error $E(T)-E$ against $T$ on a log scale for the MP2, CCSD, (T) and CCSD(T)
correlation energies of an example for which the canonical values are available (educt number 20 of the ISOL24 set
computed using a cc-pVTZ basis). Lines of best fit using $T$=$10^{-6}$--$10^{-8}$ have been computed and the $\alpha$ values are given in the legend. The behaviour shown for this example is typical of that seen across all the molecules in the ISOL24 and MOBH35 test sets.

In agreement with our previous findings, the PNO-MP2 truncation error follows the $E(T)  = E + A T^\alpha$ error model very closely,
with $\alpha=0.44$ in this case. The PNO-CCSD truncation error, on the other hand, deviates significantly from this error model, and smaller than expected errors are obtained for loose PNO thesholds. We have observed similar behaviour in the PNO truncation error for PNO-LCCD energies (linear CCD or CEPA0\cite{Ahlrichs:1987}). The key contributions to the LCCD and CCSD amplitude equations are the ring and ladder MP3 terms, which are large but have opposite sign. These converge at different rates with PNO truncation, with the ladder terms converging more slowly than the ring terms.\cite{Kohn:JCP113-174117} The ladder terms act to reduce the amplitudes and the correlation energy, whereas the ring terms act to increase the correlation energy. The slow convergence with PNO threshold of the ladder terms causes looser thresholds to have larger correlation energies than would be the case if all contributions converged at the same rate. Extrapolation of PNO-CCSD energies to the CPS limit using simple one-component error models will therefore have limited success.

We turn now to the truncation error for the (T) energy. This depends on the triple natural orbital (TNO) truncation threshold, which is set to be equal to the PNO threshold. We find that this contribution does follow the simple $E(T)  = E + A T^\alpha$ error model. In fact, the error in the (T) energy has two sources: the TNO truncation error; and the error in the T2 amplitudes used to compute the (T) energy. The error in the (T) energy is directly proportional to the TNO occupation number threshold in the same way that the error in the MP2 energy is proportional to the PNO occupation number threshold, which explains the near linearity of the log-log plot. The slight deviation from the ideal error model is a result of the error in the T2 amplitudes, and follows the trend observed for CCSD. Since the TNO error is the dominant contribution to the total error in the PNO-CCSD(T) energies, extrapolation of PNO-CCSD(T) energies to the CPS limit using simple error models is expected to be successful. If in the future the error in the (T) energy is reduced through improved TNO construction, then extrapolation of PNO-CCSD(T) to the CPS limit will become more challenging due to the increase importance of the CCSD contributions.

\begin{figure}[!tbp]
   \begin{center}
   \includegraphics[width=0.95\columnwidth]{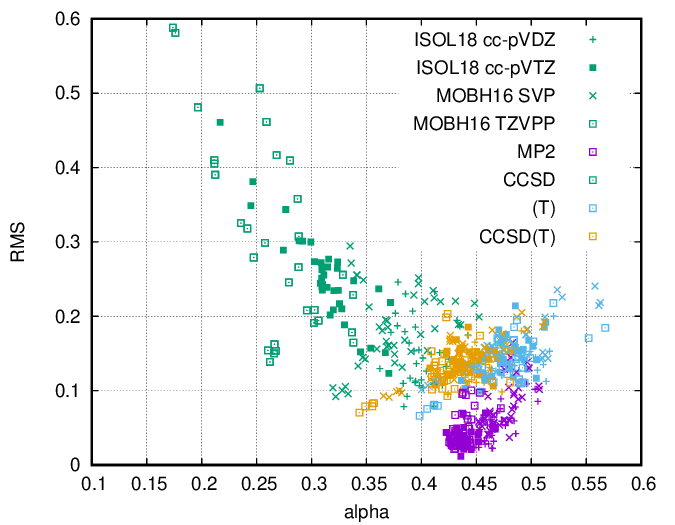}
   \end{center}
\caption{\label{fig:alpha} Values of $\alpha$ in line of best fit to $\log_{10}(E(T)-E)  = \log_{10} A + \alpha \log_{10} T$ against RMS deviation, for PNO truncation errors in MP2, CCSD and CCSD(T) energies.}
\end{figure}

For each of the molecules in our data sets where we were able to compute the canonical energies, we have perfomed a linear fit to the PNO truncation data using $\log_{10}(E(T)-E)  = \log_{10} A + \alpha \log_{10} T$. In Fig \ref{fig:alpha} we present a scatter plot of the obtained $\alpha$ against the root mean squared deviation of the data from the model. We used values of $T=10^{-5}$--$10^{-9}$ for the fits. The data is consistent with the PNO convergence shown for educt number 20 in Fig \ref{fig:example}. 
The low RMS deviations for the MP2 data indicate that the PNO-MP2 truncation follows the error model closely, and the exponent $\alpha$ is just below the ideal value of $0.5$ and is only weakly dependent on system and basis set. 

The CCSD data, however, has large RMS deviations from the model. $\alpha$ values ranging from 0.1--0.5 are obtained, reflecting varying levels of cancellation of the ring and ladder terms. The deviations are larger for the triple-zeta basis sets than the double-zeta sets, but no obvious difference is seen when contrasting the MOBH16 and ISOL18 sets. The (T) data does follow the simple error model, with modest deviations from the ideal value of $\alpha=0.5$.

\begin{figure}[!tbp]
   \begin{center}
   \includegraphics[width=0.95\columnwidth]{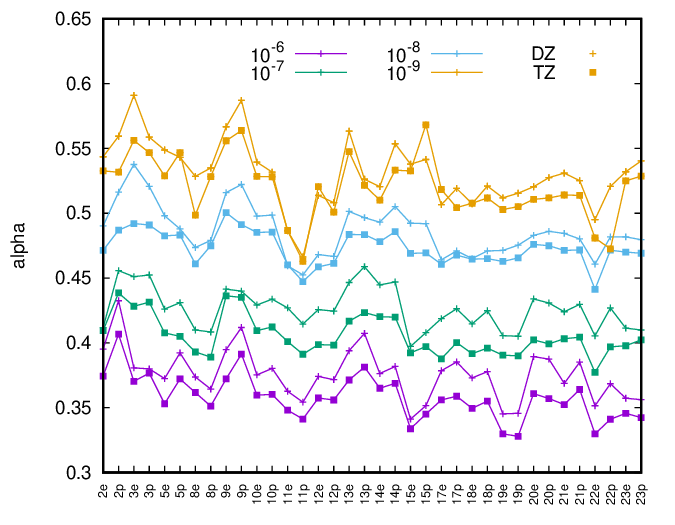}
   \end{center}
\caption{\label{fig:alpha_t} Values of $\alpha(T)$ in Eq~\ref{eq:alphat} for PNO-CCSD(T) energies of molecules in the ISOL18 set}
\end{figure}

The three-point extrapolation scheme we introduced in Ref. \onlinecite{Sorathia:JCP153-174112} determines the effective exponent $\alpha$ on a case-by-case basis from the energy convergence. For this to be accurate, the effective exponent $\alpha$ must be approximately constant over the range of $T$ used to perform the extrapolation. Given the canonical
limit $E$, the value of $\alpha$ corresponding to two thresholds $T$ and $10T$ is 
\begin{align}
\alpha(T) = \log_{10} \left( \frac{E(10T) - E}{E(T) - E} \right)
\label{eq:alphat}
\end{align}
In Fig \ref{fig:alpha_t} we display $\alpha$ for $T=10^{-6}$, $10^{-7}$, $10^{-8}$ and $10^{-9}$ for 
CCSD(T) energies of the molecules of the ISOL18 set for which we have canonical energies. Evidently,
$\alpha$ varies considerably with $T$ and the variation with $T$ is larger than the variation between 
molecules and or between basis sets. This explains why our attempts to apply the three-point extrapolation formula
to PNO-CCSD(T) energies was unsuccessful and why it is more effective to fix the exponent $\alpha$ close to the
ideal value of 0.5 and perform a two-point extrapolation.

\subsection{Error model $E(T)  = E + A(T) T^{1/2}$}

\begin{figure}[!tbp]
   \begin{center}
   \includegraphics[width=0.95\columnwidth]{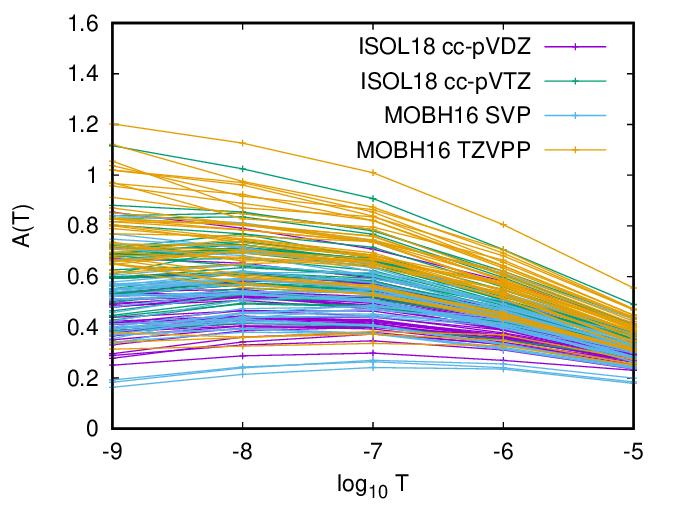}
   \end{center}
\caption{\label{fig:a} Values of $A(T)$ in $E(T)  = E + A(T) T^{1/2}$ for PNO-CCSD(T) energies in mE$_h$ per valence electron}
\end{figure}

If we fix the exponent $\alpha$ at the ideal value of 0.5, then the PNO truncation error can be written without
loss of generality as $E(T)  = E + A(T) T^{1/2}$. Two-point extraplation assumes that the positive prefactor $A$ 
is constant and will be accurate if $A(T)$ is approximately independent of $T$.
Applying the two-point extrapolation formula, we obtain
\begin{align}
\tilde E &= \frac{ T_2^{1/2} E(T_1) - T_1^{1/2} E(T_2) }{T_2^{1/2} - T_1^{1/2}}  \\
         &= E + \frac{T_1^{1/2}T_2^{1/2}}{T_2^{1/2} - T_1^{1/2}} (A(T_1) - A(T_2))
\end{align}
If $A$ increases with $T$, then the extrapolation predicts energies below the canonical limit, whereas if
$A$ decreases with $T$, the correlation energy is underestimated. In Fig~\ref{fig:a} we plot $A(T)$ for the molecules
in our test sets where we have the canonical energies. The prefactor $A$ is proprotional to the the number of 
correlated electrons in the same way as the total correlation energy and we therefore use units of mE$_h$ per valence
electron for $A$. The magnitude of $A$ reflects how strongly correlated the electrons are. $A$ is also greater for larger AO
basis sets since more of the correlation energy is recovered. Although the prefactor $A$ is not constant
as a function of $T$, for most molecules the variation is small, particularly in the range $T=10^{-6}$--$10^{-8}$, and we expect the two-point extrapolation to perform well.

\begin{table*}[!tbp]
\caption{PNO-CCSD(T) truncation error statistics for $T=10^{-6}$--$10^{-9}$ in kcal/mol with and without two-point CPS extrapolation}
\label{tab:err}
\begin{tabular*}{0.95\textwidth}{@{\extracolsep{\fill}} lllrrrrrrrr}
\hline\hline
Testset &  Basis     & Error   &$ 10^{-6} $&$ (5,6)  $&$ 10^{-7}$&$ (6,7)  $&$ 10^{-8}$&$ (7,8)  $&$ 10^{-9}$&$ (8,9) $\\
\hline
ISOL18  & cc-pVDZ    & AV      &$  -0.35  $&$ -0.30  $&$ -0.19  $&$ -0.11  $&$ -0.07  $&$ -0.02  $&$ -0.02  $&$ -0.00 $\\
        &            & RMS     &$   0.98  $&$  0.66  $&$  0.47  $&$  0.26  $&$  0.18  $&$  0.06  $&$  0.07  $&$  0.03 $\\
        &            & MAX     &$   2.39  $&$ -1.51  $&$ -1.14  $&$ -0.68  $&$ -0.43  $&$ -0.21  $&$ -0.18  $&$ -0.06 $\\
        & cc-pVTZ    & AV      &$  -0.40  $&$ -0.33  $&$ -0.21  $&$ -0.12  $&$ -0.09  $&$ -0.04  $&$ -0.03  $&$  0.00 $\\
        &            & RMS     &$   1.12  $&$  0.78  $&$  0.55  $&$  0.31  $&$  0.23  $&$  0.11  $&$  0.08  $&$  0.04 $\\
        &            & MAX     &$  -2.64  $&$ -1.86  $&$ -1.29  $&$ -0.92  $&$ -0.65  $&$ -0.42  $&$  0.22  $&$  0.09 $\\
        & (DT)       & AV      &$  -0.42  $&$ -0.34  $&$ -0.22  $&$ -0.12  $&$ -0.10  $&$ -0.04  $&$ -0.03  $&$  0.00 $\\
        &            & RMS     &$   1.19  $&$  0.83  $&$  0.58  $&$  0.34  $&$  0.26  $&$  0.14  $&$  0.09  $&$  0.05 $\\
        &            & MAX     &$  -2.85  $&$ -2.01  $&$ -1.36  $&$ -1.01  $&$ -0.75  $&$ -0.52  $&$  0.24  $&$ -0.11 $\\
%MOBH35  & def2-SVP   & AV      &$   0.60  $&$  0.42  $&$  0.29  $&$  0.14  $&$  0.12  $&$  0.05  $&$  0.05  $&$  0.02 $\\
%        &            & RMS     &$   1.50  $&$  1.07  $&$  0.84  $&$  0.58  $&$  0.44  $&$  0.28  $&$  0.23  $&$  0.14 $\\
%        &            & MAX     &$  -5.34  $&$ -3.73  $&$ -3.16  $&$ -2.15  $&$ -1.59  $&$  1.05  $&$  0.76  $&$  0.48 $\\
MOBH16  & def2-SVP   & AV      &$   0.28  $&$  0.16  $&$  0.10  $&$  0.02  $&$  0.03  $&$ -0.01  $&$  0.00  $&$ -0.01 $\\
        &            & RMS     &$   0.70  $&$  0.48  $&$  0.28  $&$  0.15  $&$  0.13  $&$  0.08  $&$  0.06  $&$  0.03 $\\
        &            & MAX     &$   1.65  $&$  1.38  $&$  0.67  $&$ -0.54  $&$  0.33  $&$ -0.29  $&$  0.14  $&$ -0.08 $\\
% JM     & def2-SVP   & AV      &$   0.27  $&$  0.15  $&$  0.09  $&$  0.00  $&$  0.01  $&$ -0.02  $&$ -0.01  $&$ -0.02 $\\
%        &            & RMS     &$   0.71  $&$  0.51  $&$  0.30  $&$  0.18  $&$  0.15  $&$  0.11  $&$  0.10  $&$  0.09 $\\
%        &            & MAX     &$  -1.75  $&$  1.61  $&$  0.78  $&$ -0.49  $&$  0.44  $&$  0.29  $&$  0.29  $&$  0.23 $\\
        & def2-TZVPP & AV      &$   0.30  $&$  0.15  $&$  0.12  $&$  0.03  $&$  0.03  $&$ -0.00  $&$  0.01  $&$ -0.00 $\\
        &            & RMS     &$   0.87  $&$  0.52  $&$  0.37  $&$  0.20  $&$  0.16  $&$  0.10  $&$  0.07  $&$  0.05 $\\
        &            & MAX     &$  -2.33  $&$  1.22  $&$ -0.89  $&$ -0.54  $&$  0.40  $&$ -0.41  $&$  0.20  $&$ -0.12 $\\
        & (ST)       & AV      &$   0.31  $&$  0.14  $&$  0.13  $&$  0.04  $&$  0.04  $&$ -0.00  $&$  0.01  $&$  0.00 $\\
        &            & RMS     &$   0.95  $&$  0.56  $&$  0.40  $&$  0.22  $&$  0.17  $&$  0.11  $&$  0.08  $&$  0.06 $\\
        &            & MAX     &$  -2.63  $&$ -1.21  $&$ -1.01  $&$ -0.54  $&$  0.42  $&$ -0.46  $&$  0.23  $&$ -0.14 $\\
\hline\hline
\end{tabular*}
\end{table*}

In Table \ref{tab:err} we report average (AV), root mean squared (RMS) and maximum (MAX) deviations 
from the CPS limit for PNO-CCSD(T) energies
for the ISOL18 isomerisation energies and the MOBH16 barrier heights.
Values with PNO threshold $T=10^{-6}$--$10^{-9}$ are presented, together with two-point CPS extrapolation, where for example
(6,7) denotes extrapolation using $T=10^{-6}$ and $T=10^{-7}$. The two-point CPS extrapolation of Eq.~\ref{eq:t2} is used with
$\alpha=0.5$, which corresponds to $F=1.462$ in Eq.~\ref{eq:t2f}. 

CPS extrapolation reduces the PNO error by approximately 
a factor of two, which is almost equivalent to reducing the PNO threshold by one order of magnitude. This observation holds
for both test sets and all basis sets used. RMS errors using the
default threshold of $T=10^{-7}$ are half a kcal/mol, with outliers around 1.5~kcal/mol.
The default threshold is thus not sufficient to ensure that PNO truncation errors in energy differences are smaller than 
1~kcal/mol target of chemical accuracy. CPS (6,7) extrapolation improves this situation markedly, although the outliers
are still around 1~kcal/mol. To ensure that PNO truncation errors are within chemical accuracy, 
it is necessary to use the very tight treshold of $T=10^{-8}$. With (7,8) CPS extrapolation, the maximum truncation 
errors for our data sets are 0.5 kcal/mol.

Table \ref{tab:err} also includes the corresponding values for CBS extrapolation where we use PNO-CCSD(T) energies with
two basis sets to extrapolate to the complete basis set limit. For simplicity, we use Helgaker's two-point approach\cite{HelgakerJCP:106-31} with
Cardinal number $2$ for the def2-SVP and cc-pVDZ basis sets and $3$ for the def2-TZVPP and cc-pVTZ basis sets. We observe that
the PNO truncation error increases with basis size, and is magnified slightly when performing CBS extrapolation due to
the propagation of errors. It is therefore even more important to use tight PNO thresholds and CPS extrapolation.
This underlines the conclusions of our previous work.\cite{Sorathia:JCP153-174112}

The cost of a PNO-CCSD(T) calculation increases by a factor of 2-3 with every ten-fold decrease of $T$, 
and increases by a factor of 2-3 with every increment in the Cardinal number of the AO basis.
Performing PNO-CCSD(T) calculations with large basis sets and tight thresholds is expensive and can exceed
the limits of commonly available disk and memory resources.
F12 explicitly correlated methods\cite{Hattig:CR112-4} are a good solution to this computational bottleneck. 
It is, however, very useful to be able to access the basis set limit using regular methods. 

One approach to reducing the PNO truncation error of
PNO-CCSD(T) calculations with a large basis is to estimate the error using a smaller basis set or a lower cost method
and add a correction term.\cite{Neese:JCP130-114108,Pogrebetsky:JCTC19-4023}
This assumes that the PNO truncation error is approximately constant across methods and 
basis sets, but, as we have previously noted, the prefactor $A(T)$ in fact has a significant basis set dependence. It has
an even larger variation with correlation method, since different proportions of the correlation energy are recovered.

\begin{figure}[!tbp]
   \begin{center}
   \includegraphics[width=0.95\columnwidth]{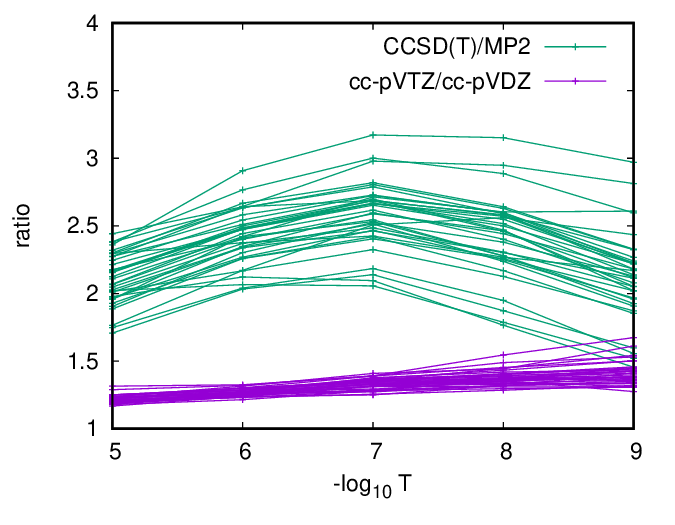}
   \end{center}
\caption{\label{fig:ratio} Ratios of $A(T)$ between the cc-pVTZ and cc-pVDZ basis sets and between the CCSD(T) and MP2 methods for
the ISOL18 set.}
\end{figure}

However, we find that the ratio between the $A(T)$ for different 
basis sets is only weakly dependent on $T$. To a lesser extent, the variation in the ratio
between $A(T)$ for different methods is also relatively small.
This is seen from Fig.~\ref{fig:ratio} where we plot the ratio between $A(T)$ for the cc-pVDZ and cc-pVTZ basis sets for the 
molecules of the ISOL18 set, together with the ratio between $A(T)$ for the CCSD(T) and MP2
correlation energies in the cc-pVTZ basis.
We can therefore accurately estimate the scaling factor that
relates the PNO truncation error for one method or basis set with another
\begin{align}
E_X(T) &= E_X + f(T) A_Y(T) T^{1/2} \\
f(T) &= \frac{A_X(T)}{A_Y(T)} \approx \frac{E_X(T_2) - E_X(T_1)}{E_Y(T_2) - E_Y(T_1)}
\end{align}
Here $X$ denotes an expensive method and basis set combination, and $Y$ denotes a less demanding approach.
Since $f$ is only weakly dependent on $T$, it can be computed using relatively loose PNO thresholds with low cost.
Applying two-point extrapolation leads to the following simple formula for the CPS limit for method $X$
\begin{align}
E_X &\approx E_X(T_3) + f F ( E_Y(T_4) - E_Y(T_3) )  \label{eq:f} \\
F &= \frac{T_3^{1/2}}{T_3^{1/2} - T_4^{1/2}}, \quad
f = \frac{E_X(T_2) - E_X(T_1)}{E_Y(T_2) - E_Y(T_1)}
\end{align}
The PNO truncation thresholds should be chosen such that $T_1 > T_2 \ge T_3 > T_4$. If $T_2=T_3$,
then five calculations are required in total.

\begin{table}[!tbp]
\caption{PNO truncation error statistics for CCSD(T)/cc-pVTZ isomerisation energies of the ISOL18 set using Eq~\ref{eq:f}}
\label{tab:f}
%\begin{tabular}{0.95\textwidth}{@{\extracolsep{\fill}} llrrr}
\begin{tabular}{llrrr}
\hline\hline
$Y$                  & Error   &$ (5,6,7) $&$ (6,7,8) $&$ (7,8,9) $\\
\hline
CCSD(T)/cc-pVTZ      & AV      &$  -0.12  $&$  -0.03  $&$   0.00  $\\
                     & RMS     &$   0.31  $&$   0.11  $&$   0.04  $\\
                     & MAX     &$  -0.92  $&$  -0.42  $&$   0.09  $\\
CCSD(T)/cc-pVDZ      & AV      &$  -0.09  $&$  -0.00  $&$  -0.02  $\\
                     & RMS     &$   0.32  $&$   0.13  $&$   0.07  $\\
                     & MAX     &$  -0.86  $&$  -0.35  $&$  -0.24  $\\
MP2/cc-pVTZ          & AV      &$  -0.19  $&$  -0.11  $&$  -0.01  $\\
                     & RMS     &$   0.83  $&$   0.23  $&$   0.14  $\\
                     & MAX     &$   2.31  $&$  -0.71  $&$  -0.42  $\\
MP2/cc-pVDZ          & AV      &$  -0.14  $&$  -0.11  $&$   0.01  $\\
                     & RMS     &$   0.79  $&$   0.30  $&$   0.12  $\\
                     & MAX     &$   1.76  $&$  -0.70  $&$  -0.43  $\\
\hline\hline
\end{tabular}
\end{table}

We have tested the accuracy of Eq.~\ref{eq:f} for the CCSD(T)/cc-pVTZ isomerisation energies of the ISOL18 set.
In Table~\ref{tab:f} we report deviations from the CPS limit for different choices of method $Y$. The notation
(5,6,7) refers to $T_1=10^{-5}$, $T_2=T_3=10^{-6}$, $T_4=10^{-7}$, etc. For comparison,
the values obtained with $Y$=CCSD(T)/cc-pVTZ are also listed, which are identical to simply applying
Eq.~\ref{eq:t2f} with $T_3$ and $T_4$.

Comparing the CCSD(T)/cc-pVTZ and CCSD(T)/cc-pVDZ results, we see that the accuracy is very similar. Therefore,
to obtain (7,8) quality PNO-CCSD(T)/cc-pVTZ values, it is not necessary to perform PNO-CCSD(T)/cc-pVTZ 
calculations with $T=10^{-7}$ and $10^{-8}$. Instead, $T=10^{-6}$ and $10^{-7}$ are required, together with
the significantly cheaper $T=10^{-6}$, $10^{-7}$ and $10^{-8}$ calculations using the smaller cc-pVDZ basis.

Comparing the CCSD(T)/cc-pVTZ and MP2/cc-pVTZ results, we see that there is a marked reduction in accuracy when
using MP2. The variation of the factor $f$ in Eq~\ref{eq:f} with $T$ is greater and less systematic when 
changing method than changing basis, and the uncertainty in the extrapolated energies is correspondingly larger.
Since PNO-MP2 calculations are much less expensive than PNO-CCSD(T) calculations, it is nevertheless potentially
worthwhile to use $(6,7,8)$ with MP2 since the results are a slight improvement over $(6,7)$ without the MP2 correction.
Reducing both the method and the basis set introduces too large errors and is not recommended.

In Table~\ref{tab:13} we present PNO truncation errors for reaction 13 of the MOBH35 set, as an example of a
system with large static correlation and slow PNO convergence. We compare different schemes for reducing
the PNO truncation error of PNO-CCSD(T)/def2-TZVPP energies: no extraplation; adding an MP2 correction $\Delta$ as
advocated by Kubas;\cite{Pogrebetsky:JCTC19-4023} scaled extrapolation using the def2-SVP basis; and 
straightforward two-point extrapolation.
For each method we report the sum of the wall times taken to esimate the canonical energy of the reactant. All
timings include the HF calculation, which took 12 minutes using density fitting. Although adding
a correction $\Delta = E_\text{MP2} - E_\text{PNO-MP2}$ does reduce the PNO truncation error for
loose thesholds, with minimal additional expense, it is rather ineffective for tight thresholds. The most
cost-effective way to ensure that the canonical result is recovered is to use the (6,7,8) scaled
extrapolation scheme, which avoids the expense of performing a PNO-CCSD(T)/def2-TZVPP calculation with
the very tight threshold of $10^{-8}$.

\begin{table}[!tbp]
\caption{Timings and PNO truncation errors in kcal/mol of CCSD(T)/def2-TZVPP barrier heights for MOBH35 reaction 13 using
PNO thresholds $10^{-6}$-$10^{-8}$ and with extrapolation and correction schemes.}
\label{tab:13}
\begin{tabular*}{0.95\columnwidth}{@{\extracolsep{\fill}} lrrr}
\hline\hline
Scheme      & Forward & Reverse& Minutes \\
\hline
6           &$  5.5  $&$ 3.1  $&$  92  $\\
6+$\Delta$  &$  3.7  $&$ 2.2  $&$  93  $\\
(5,6)       &$  3.7  $&$ 1.9  $&$ 144  $\\
7           &$  2.5  $&$ 1.4  $&$ 184  $\\
7+$\Delta$  &$  1.8  $&$ 1.1  $&$ 185  $\\
(5,6,7)     &$  1.0  $&$ 1.0  $&$ 213  $\\
(6,7)       &$  1.1  $&$ 0.6  $&$ 265  $\\
8           &$  0.9  $&$ 0.5  $&$ 554  $\\
8+$\Delta$  &$  0.6  $&$ 0.4  $&$ 555  $\\
(6,7,8)     &$  0.1  $&$ 0.3  $&$ 398  $\\
(7,8)       &$  0.1  $&$ 0.1  $&$ 726  $\\
\hline\hline
\end{tabular*}
\end{table}

\section{Benchmark data}

\subsection{MOBH35}

\begin{table*}[!tbp]
\caption{Best estimates for CCSD(T) barrier heights for the MOBH35 test set in kcal/mol}
\label{tab:mobh35}
\begin{tabular*}{0.95\textwidth}{@{\extracolsep{\fill}} lrrrrrrrrrr}
\hline\hline
        & \multicolumn{5}{c}{Forward} & \multicolumn{5}{c}{Reverse} \\
rxn     &  SVP    &  TZVPP  &   (ST)   &   QZVPP  &    (TQ)   &     SVP    &    TZVPP  &   (ST)  &    QZVPP  &    (TQ) \\
\hline
1       &$ 27.06 $&$ 26.67 $&$  26.43 $&$  26.26 $&$  25.94  $&$   14.02  $&$  13.91  $&$ 13.87 $&$  14.45  $&$ 14.75 $ \\
2       &$  5.62 $&$  5.91 $&$   5.62 $&$   5.84 $&$   5.77  $&$   25.10  $&$  22.25  $&$ 21.87 $&$  22.34  $&$ 22.43 $ \\
3       &$  0.95 $&$  1.03 $&$   0.90 $&$   1.01 $&$   1.01  $&$   27.07  $&$  26.14  $&$ 26.80 $&$  27.09  $&$ 27.69 $ \\
4       &$  2.35 $&$  1.54 $&$   0.98 $&$   1.45 $&$   1.38  $&$    8.60  $&$   7.87  $&$  8.34 $&$   8.43  $&$  8.82 $ \\
5       &$  4.43 $&$  4.69 $&$   4.28 $&$   4.94 $&$   5.03  $&$   22.09  $&$  22.73  $&$ 22.60 $&$  22.64  $&$ 22.68 $ \\
6       &$ 13.48 $&$ 15.60 $&$  15.81 $&$  15.84 $&$  15.98  $&$   13.46  $&$  14.82  $&$ 14.32 $&$  15.02  $&$ 14.98 $ \\
7       &$ 26.62 $&$ 27.70 $&$  28.01 $&$  27.72 $&$  27.80  $&$   18.26  $&$  18.91  $&$ 18.29 $&$  18.91  $&$ 18.85 $ \\
8$^a$   &$ 37.28 $&$ 35.65 $&$  35.69 $&$  36.01 $&$  35.70  $&$   32.77  $&$  32.54  $&$ 32.47 $&$  33.46  $&$ 33.70 $ \\
9$^a$   &$ 28.97 $&$ 28.08 $&$  28.50 $&$  29.42 $&$  30.01  $&$   14.90  $&$  12.07  $&$ 10.66 $&$  11.09  $&$ 10.82 $ \\
10      &$ -3.52 $&$ -3.82 $&$  -3.79 $&$  -4.16 $&$  -4.43  $&$    9.59  $&$   8.95  $&$  7.41 $&$   8.03  $&$  7.56 $ \\
11$^a$  &$ 29.89 $&$ 30.05 $&$  29.91 $&$  29.59 $&$  29.39  $&$   84.13  $&$  83.17  $&$ 83.10 $&$  82.95  $&$ 82.54 $ \\
12      &$  5.68 $&$  5.37 $&$   5.46 $&$   5.31 $&$   5.28  $&$   36.95  $&$  37.38  $&$ 38.33 $&$  37.17  $&$ 37.19 $ \\
13      &$ 18.85 $&$ 20.69 $&$  21.37 $&$  20.88 $&$  21.24  $&$   48.37  $&$  48.59  $&$ 48.89 $&$  48.52  $&$ 48.42 $ \\
14      &$ 10.20 $&$ 10.26 $&$  10.26 $&$  10.31 $&$  10.35  $&$   13.33  $&$  14.43  $&$ 13.95 $&$  14.89  $&$ 14.99 $ \\
15      &$ 23.84 $&$ 20.77 $&$  19.77 $&$  20.44 $&$  20.14  $&$   74.62  $&$  74.91  $&$ 74.55 $&$  75.96  $&$ 76.51 $ \\
16      &$ 37.24 $&$ 35.09 $&$  34.50 $&$  35.04 $&$  34.89  $&$   55.45  $&$  53.61  $&$ 53.84 $&$  53.93  $&$ 54.37 $ \\
17$^a$  &$ 24.22 $&$ 22.84 $&$  24.20 $&$  21.28 $&$  20.54  $&$   29.94  $&$  28.35  $&$ 27.54 $&$  30.94  $&$ 31.87 $ \\
18$^a$  &$ 25.53 $&$ 26.12 $&$  27.79 $&$  24.94 $&$  24.44  $&$   27.93  $&$  28.35  $&$ 27.83 $&$  31.70  $&$ 33.09 $ \\
19$^a$  &$ 11.05 $&$ 11.55 $&$  13.09 $&$  10.96 $&$  10.58  $&$   30.36  $&$  28.80  $&$ 27.98 $&$  31.69  $&$ 32.85 $ \\
20$^a$  &$  7.60 $&$ 10.49 $&$  12.28 $&$  10.29 $&$  10.06  $&$   28.15  $&$  28.71  $&$ 28.20 $&$  32.18  $&$ 33.63 $ \\
21      &$ 11.06 $&$  8.14 $&$   7.87 $&$   8.44 $&$   8.70  $&$   11.06  $&$   8.15  $&$  7.88 $&$   8.44  $&$  8.69 $ \\
22      &$ 14.90 $&$ 14.42 $&$  13.64 $&$  14.39 $&$  14.37  $&$   30.83  $&$  27.25  $&$ 26.42 $&$  27.66  $&$ 28.00 $ \\
23      &$ 29.40 $&$ 29.98 $&$  28.81 $&$  29.90 $&$  29.85  $&$   20.82  $&$  20.45  $&$ 20.12 $&$  20.64  $&$ 20.79 $ \\
24$^a$  &$  1.08 $&$  2.44 $&$   2.35 $&$   2.71 $&$   2.85  $&$   18.61  $&$  17.03  $&$ 17.34 $&$  16.61  $&$ 16.54 $ \\
25$^a$  &$  1.52 $&$  2.78 $&$   2.60 $&$   3.11 $&$   3.23  $&$   14.58  $&$  12.90  $&$ 13.08 $&$  12.91  $&$ 13.08 $ \\
26      &$ 21.79 $&$ 25.07 $&$  25.94 $&$  25.25 $&$  25.39  $&$   -0.07  $&$   0.13  $&$  0.22 $&$   0.14  $&$  0.17 $ \\
27      &$ 16.11 $&$ 14.09 $&$  14.00 $&$  13.87 $&$  13.84  $&$    1.29  $&$   1.88  $&$  1.75 $&$   2.20  $&$  2.42 $ \\
28      &$ 32.00 $&$ 30.63 $&$  30.50 $&$  30.21 $&$  29.86  $&$   16.85  $&$  15.94  $&$ 15.41 $&$  15.84  $&$ 15.81 $ \\
29      &$ 15.76 $&$ 15.30 $&$  15.05 $&$  14.91 $&$  14.69  $&$   33.87  $&$  32.02  $&$ 30.74 $&$  31.35  $&$ 30.88 $ \\
30      &$ 10.94 $&$ 10.07 $&$   9.94 $&$   9.81 $&$   9.64  $&$   19.89  $&$  17.10  $&$ 16.29 $&$  17.19  $&$ 17.15 $ \\
31      &$  2.37 $&$  3.41 $&$   4.00 $&$   3.26 $&$   3.07  $&$   12.31  $&$  13.35  $&$ 13.03 $&$  13.01  $&$ 12.79 $ \\
32      &$ 23.56 $&$ 20.43 $&$  20.31 $&$  20.01 $&$  19.87  $&$   58.31  $&$  62.06  $&$ 62.67 $&$  63.31  $&$ 64.08 $ \\
33      &$  2.76 $&$  1.21 $&$   0.70 $&$   1.13 $&$   0.99  $&$   10.00  $&$   8.24  $&$  8.21 $&$   8.10  $&$  8.02 $ \\
34      &$ 28.85 $&$ 29.82 $&$  29.56 $&$  29.11 $&$  28.66  $&$    4.38  $&$   3.54  $&$  3.49 $&$   3.24  $&$  2.99 $ \\
35      &$ 15.00 $&$ 16.68 $&$  16.51 $&$  17.54 $&$  17.99  $&$   -3.74  $&$  -2.42  $&$ -2.24 $&$  -2.08  $&$ -1.86 $ \\
\hline\hline
\end{tabular*}\\
$^a$ def2-QZVPP values are computed using the (6,7,8) threshold combination with $Y$=def2-TZVPP, rather than
using the (7,8,9) threshold combination with $Y$=def2-SVP.
\end{table*}

Our CPS extrapolation approach makes it possible to reliably estimate the canonical CCSD(T) energies of large systems with 
large basis sets using PNO methods, and thus extrapolate to the CBS limit without being limited by PNO truncation errors. 
In Table~\ref{tab:mobh35} we report our best estimates for the canonical CCSD(T) barrier heights of the full MOBH35 set.
Our def2-SVP values agree closely with those previously reported and we present for the first time def2-QZVPP values
for the full set, including reactions 17--20, which were ommited from the work of Semidalas and Martin.

To compute the CPS limit for CCSD(T)/def2-SVP, we use (8,9) extrapolation of PNO-CCSD(T) energies based on 
the $T^{1/2}$ convergence model, which has an RMS deviation from the canonical limit of under 0.1~kcal/mol.
PNO-CCSD(T)/def2-SVP calculations using $T=10^{-9}$ were possible for the full data set using the Turbomole implementation.
Disk space limitations precluded the use of $T=10^{-9}$ with the def2-TZVPP, but $T=10^{-8}$ 
was possible for all molecules. To estimate the CCSD(T)/def2-TZVPP canonical limit, we use (7,8,9) extrapolation 
\begin{align}
E_{T} &\approx E_{T}(10^{-8}) + f F ( E_S(10^{-9}) - E_S(10^{-8}) ) \\
F &= \frac{10^{1/2}}{10^{1/2} - 1}, \quad
f = \frac{E_T(10^{-8}) - E_T(10^{-7})}{E_S(10^{-8}) - E_S(10^{-7})}
\end{align}
It was also possible to compute PNO-CCSD(T)/def2-QZVPP values for all but the largest molecules and we also used the
(7,8,9) extrpaolation with $Y$=def2-SVP. For the largest molecules, we used (6,7,8) extrapolation with $Y$=def2-TZVPP
\begin{align}
E_{Q} &\approx E_{Q}(10^{-7}) + f F ( E_T(10^{-8}) - E_T(10^{-7}) ) \\
F &= \frac{10^{1/2}}{10^{1/2} - 1}, \quad
f = \frac{E_Q(10^{-7}) - E_Q(10^{-6})}{E_T(10^{-7}) - E_T(10^{-6})}
\end{align}
Our def2-SVP values do not agree perfectly with the subset of canonical values reported by Semidalas and Martin.
Their data is based on HF energies computed using density fitting, whereas we did not employ this approximation in our HF calculations, and the difference in the HF energies and the resulting change in the correlation energies amounts to 0.1~kcal/mol deviations on average. Reactions 5, 6, 12, 24, 25, 26, 31, 32 and 35 have deviations between 0.1-0.3~kcal/mol, which is not untypical.\cite{Tew:JCP148-011102} In addition, Table~\ref{tab:mobh35} reports (8,9) extrapolated CPS values rather than canonical values. Although these are within 0.1~kcal/mol of the canonical barrier heights for the MOBH16 set, this does not contain reactions 8, 9 and 13, which are more strongly correlated and converge more slowly with PNO threshold. Residual CPS errors of 0.4~kcal/mol may remain for these reactions. A conservative error bar of 0.2~kcal/mol should be placed on the PNO estimates for the canonical values except for reactions 8, 9 and 13, which may have errors of 0.5~kcal/mol.

The primary difference in our best CBS estimate to that of Semidalas and Martin is that we have performed a (TQ) extrapolation
for both the CCSD and (T) correlation energies. Semidalas and Martin did not compute the (T) contribution using the 
def2-QZVPP basis and instead used a (ST) extrapolation for the (T) energy. Nevertheless, our barrier heights differ by less than 0.5~kcal/mol to their values for all reactions except for 8, 9 and 13. For these more strongly correlated systems, our values differ by up to 2.5~kcal/mol and in fact lie between those of Semidalas and Martin\cite{Semidalas:JCTC18-883} and the original values reported by Iron and Janes.\cite{Iron:JPCA123-3761}

\subsection{ISOL24}

\begin{table}[!tbp]
\caption{Best estimates for CCSD(T) isomerisation energies for the ISOL24 test set in kcal/mol}
\label{tab:isol24}
\begin{tabular*}{0.95\columnwidth}{@{\extracolsep{\fill}} lrrrrrr}
\hline\hline
iso     &  DZ   &  TZ    &  (DT) &   QZ   & (TQ)  & F12$^a$ \\
\hline
1       &$ 67.35 $&$  70.22 $&$ 70.71 $&$  71.24 $&$ 71.55 $&$  71.53 $\\
2       &$ 40.78 $&$  39.14 $&$ 39.90 $&$  38.59 $&$ 38.68 $&$  38.13 $\\
3       &$  4.29 $&$   8.76 $&$  9.70 $&$   9.67 $&$ 10.21 $&$  10.29 $\\
4       &$ 75.61 $&$  71.55 $&$ 73.90 $&$  70.17 $&$ 69.89 $&$  69.03 $\\
5       &$ 34.76 $&$  32.51 $&$ 32.92 $&$  32.65 $&$ 33.02 $&$  32.85 $\\
6       &$ 18.62 $&$  22.41 $&$ 23.44 $&$  23.47 $&$ 24.07 $&$  24.14 $\\
7       &$ 18.90 $&$  17.92 $&$ 18.46 $&$  17.83 $&$ 17.90 $&$  17.72 $\\
8       &$ 23.04 $&$  24.00 $&$ 23.13 $&$  24.24 $&$ 24.01 $&$  23.93 $\\
9       &$ 22.77 $&$  21.68 $&$ 21.63 $&$  21.36 $&$ 21.13 $&$  21.10 $\\
10      &$  6.07 $&$   6.14 $&$  6.37 $&$   6.43 $&$  6.47 $&$   6.26 $\\
11      &$ 35.66 $&$  35.45 $&$ 35.19 $&$  36.18 $&$ 36.65 $&$  36.80 $\\
12      &$  0.68 $&$   0.43 $&$  0.43 $&$   0.42 $&$  0.48 $&$   0.42 $\\
13      &$ 29.00 $&$  32.25 $&$ 32.84 $&$  33.03 $&$ 33.29 $&$  33.20 $\\
14      &$  4.26 $&$   4.68 $&$  5.30 $&$   4.97 $&$  5.06 $&$   4.92 $\\
15      &$ 10.82 $&$   4.37 $&$  3.56 $&$   4.03 $&$  4.15 $&$   4.07 $\\
16      &$ 20.68 $&$  22.69 $&$ 23.61 $&$  22.76 $&$ 23.04 $&$  22.62 $\\
17      &$  9.51 $&$   9.19 $&$  9.57 $&$   9.29 $&$  9.45 $&$   9.59 $\\
18      &$ 27.06 $&$  24.78 $&$ 24.60 $&$  24.19 $&$ 23.93 $&$  23.77 $\\
19      &$ 18.77 $&$  18.41 $&$ 18.06 $&$  18.41 $&$ 18.37 $&$  18.28 $\\
20      &$  6.18 $&$   5.37 $&$  4.98 $&$   5.18 $&$  4.95 $&$   5.13 $\\
21      &$ 11.17 $&$  11.00 $&$ 11.51 $&$  11.29 $&$ 11.60 $&$  11.38 $\\
22$^b$  &$ -3.59 $&$  -0.98 $&$ -1.25 $&$   0.09 $&$  0.18 $&$   1.08 $\\
23      &$ 24.04 $&$  23.93 $&$ 24.31 $&$  23.72 $&$ 23.75 $&$  23.66 $\\
24      &$ 16.32 $&$  15.58 $&$ 15.49 $&$  15.34 $&$ 15.25 $&$  15.25 $\\
\hline\hline
\end{tabular*}\\
$^a$ PNO-LCCSD(T)-F12b/APVQZ' values taken from Ref~\onlinecite{Werner:JCTC}\\
$^b$ product and educt reversed to maintain positive sign
\end{table}

CBS data for the ISOL24 set have very recently been computed by Werner and Hansen using the
PNO-LCCSD(T)-F12b method in Molpro and a modified aug-cc-pVQZ basis set. In Table~\ref{tab:isol24}
we compare their isomerisation energies with our values computed using CBS extrapolation of
CPS extrapolated PNO-CCSD(T) energies.

We were able to compute PNO-CCSD(T) energies with $T=10^{-9}$ for the cc-pVDZ basis and 
$T=10^{-8}$ for the cc-pVTZ and cc-pVQZ basis sets for all molecules. Our best estimate
of canonical CCSD(T) energies is therefore using an (8,9) extrapolation for the cc-pVDZ basis
and a (7,8,9) extrapolation with $Y$=cc-pVDZ for the cc-pVTZ and cc-pVQZ basis sets.

Our best CBS values are from (TQ) extrapolation. The (TQ) and F12 isomerisation energies agree to within 
0.5~kcal/mol for all isomer pairs except for 22, where the difference is 0.9~kcal/mol. This level of
agreement is only slightly worse than that expected from canonical theory and these results
underline the viability of using PNO-CCSD(T) theory in CBS extrapolation, provided that the PNO truncation
error can be properly controlled.

\section{Conclusions}

Domain based PNO-CCSD(T) theory provides a low-scaling approximation to canonical CCSD(T) theory that
makes it possible to perform accurate calculations on large molecular systems. However, for 
such calculations to achieve so-called ``gold standard'' status and be used to predict
reaction enthalpies to within 1~kcal/mol of experiment, it is necessary to ensure that both the 
AO basis set truncation error and the PNO truncation error are sufficiently converged. 

The smooth convergence of the correlation energies with basis set size for canonical theories is well 
documented to follow an $E_X = E_\infty + C X^{-3}$ basis set error model with Cardinal number $X$,
and extrapolation to the CBS limit is routinely applied. In this article, we have demonstrated that the 
PNO truncation error for the CCSD(T) energy follows $E_X(T) = E_X + A_X T^{1/2}$ with PNO truncation
threshold $T$, so that the combined convergence is $E_X(T) = E_\infty + A_X T^{1/2} + C X^{-3}$.
The prefactor $A_X$ is basis set dependent, being greater for larger basis sets, is proportional to 
the number of correlated electrons, and is larger for more strongly correlated systems. 

To accurately obtain CBS CCSD(T) energies using PNO methods, the most reliable approach is to 
first eliminate the PNO truncation error through CPS extrapolation to obtain canonical energies
$E_X$ and $E_{(X+1)}$ and to perform two-point extrapolation in the usual manner. CPS extrapolation to the 
canonical limit proceeds in an analogous way to CBS extraplation and requires calculations with
two PNO thresholds; typically we chose $10T$ and $T$. We find that for systems with
moderate static correlation, extrapolation using ``tight'' PNO thresholds of $T=10^{-7}$ are not sufficient 
to ensure that the PNO truncation errors are less than 1~kcal/mol. Reliable results are, however, obtained 
for all cases in the ISOL24 and MOBH35 data sets when extrapolating using $T=10^{-7}$ and $T=10^{-8}$. 

Regarding the basis set, it is well documented\cite{mest} that
basis set errors in CCSD(T) calculations using double-zeta quality basis sets are commensurate with
the uncertainties in density functional approximations and that (TQ) extrapolation is the minimum
required to achieve ``gold standard'' results. The combination of tight PNO thresholds and large basis sets
places a heavy burden on current PNO-CCSD(T) implementations, particularly in the I/O of pair-specific
integrals stored on disk. We have found that the prefactor $A_X$ for a large basis set can be
accurately estimated using information from PNO calculations using a smaller basis set, and that the most expensive calculations
in the extrapolation proceedure can be avoided with very little loss of accuracy.

Our recommended CPS extrapolation approach to obtain $E_X$, the CCSD(T) energy in basis set with Cardinal number $X$, is to use
\begin{align}
E_X &\approx E_X(10^{-7}) + f F ( E_{X-1}(10^{-8}) - E_{X-1}(10^{-7}) )  \\
F &= \frac{10^{1/2}}{10^{1/2} - 1}, \quad
f = \frac{E_X(10^{-7}) - E_X(10^{-6})}{E_{X-1}(10^{-7}) - E_{X-1}(10^{-6})}
\end{align}

\bibliography{pnorefs}

%\newpage

%%%%%%%%%%%%%%%%%%%%%%%%%%%% APPENDIX %%%%%%%%%%%%%%%%%%%%%%%%%%
%\section*{Appendix}

\end{document}